\documentclass[useAMS,usenatbib,usegraphicx]{mn2e}

\title[Synthetic LCs of Shocked Dense Circumstellar Shells]
{
Synthetic
Light Curves of Shocked Dense Circumstellar Shells
}
\author[T. J. Moriya et al.]
{Takashi J. Moriya$^{1,2,3}$\thanks{takashi.moriya@ipmu.jp},
 Sergei I. Blinnikov$^{4,5,6,1}$,
Petr V. Baklanov$^{4,5}$, \newauthor
Elena I. Sorokina$^{6}$,
Alexander D. Dolgov$^{4,5,7}$ \\
$^{1}$
Kavli Institute for the Physics and Mathematics of the Universe,
Todai Institutes for Advanced Study,
University of Tokyo, \\ 5-1-5 Kashiwanoha, Kashiwa, Chiba 277-8583, Japan
\\
$^{2}$
Department of Astronomy, Graduate School of Science, University of
Tokyo, 7-3-1 Hongo, Bunkyo, Tokyo 113-0033, Japan \\
$^{3}$
Research Center for the Early Universe, Graduate School of Science, University of Tokyo,
7-3-1 Hongo, Bunkyo, Tokyo 113-0033, Japan\\
$^{4}$
Institute for Theoretical and Experimental Physics, Bolshaya Cheremushkinskaya 25, 117218 Moscow, Russia\\
$^{5}$
Novosibirsk State University, Novosibirsk 630090, Russia\\
$^{6}$
Sternberg Astronomical Institute, M.V.Lomonosov Moscow State University,
Universitetski pr. 13, 119992 Moscow, Russia\\
$^{7}$ 
INFN and Universit\`a degli Studi di Ferrara, I-44100 Ferrara, Italy 
}
\begin{document}


\pagerange{\pageref{firstpage}--\pageref{lastpage}} \pubyear{2013}

\maketitle

\label{firstpage}

\begin{abstract}
We numerically investigate light curves (LCs) 
 of shocked circumstellar shells
which are suggested to reproduce the observed LC of 
superluminous SN 2006gy analytically.
In the previous analytical model,
the effects of the recombination and the bolometric correction
on LCs are not taken into account.
To see the effects,
we perform numerical radiation hydrodynamic calculations
of shocked shells by using \verb|STELLA|,
which can numerically treat multigroup radiation transfer with
realistic opacities.
We show that the effects of the recombination and the bolometric
 correction are significant and the analytical model should 
be compare to the bolometric LC instead of a single band LC.
We find that shocked circumstellar shells have a rapid
LC decline initially because of the adiabatic expansion 
rather than the luminosity increase and the shocked shells fail 
to explain the LC properties of SN 2006gy.
However, our synthetic LCs are qualitatively similar to
those of superluminous SN 2003ma and SN 1988Z
and they may be related to shocked circumstellar shells.
\end{abstract}


\section{Introduction}\label{sec1}
Origins of superluminous supernovae (SLSNe) whose existence are
recognized recently (see \citealt{gal-yam2012} for a review) are still mystery.
Several mechanisms to explain their huge luminosities are suggested so far (e.g.,
\citealt{smith2007b,woosley2007,kasen2010,moriya2012b,manos2012,dexter2012,ouyed2012}).
At least for a subclass of SLSNe which is named as 'SLSN-II' in \citet{gal-yam2012},
ejecta-circumstellar medium (CSM) interaction is
likely to be a main power source because most of them show
narrow emission lines which are supposed to be related to the existence
of the dense circumstellar shell \citep[see, e.g.,][]{filippenko1997}.
SNe can be very bright at their early epochs with the dense
circumstellar shell since the kinetic energy of SN ejecta can be
efficiently converted to radiation due to the interaction.

Dense circumstellar shell causing SLSNe are likely to be dense enough to
cause the shock breakout in the shell \citep[e.g.,][]{moriya2012b}.
There are two possible locations at which shock breakout can occur
in a dense shell.
The two cases are clearly summarized in \citet{chevalier2011}.
In one case, the optical depth of the dense shell is high enough to keep
photons in the forward shock until it gets near the surface of the dense
shell.
In other words, the shock breakout occurs near the surface of the dense shell.
On the other hand, if the optical depth is low, the shock breakout
can occur within the shell well 
before the shock reaches the vicinity of the shell surface. 
In this case, photons emitted from the shock diffuse in the remaining
shell after the shock breakout. LCs from the latter cases have been numerically
investigated and they are found to be consistent with the LC of SN
2006gy \citep{moriya2012b, ginzburg2012}.

\begin{table}
\centering
\begin{minipage}{120mm}
\caption{List of initial conditions}
\label{table1}
\begin{tabular}{cccccc}
\hline
Name & $v_o$\footnote{velocity of the outermost layer of the shell}  &
$M_s$\footnote{mass of the shell} &
 $R_o$\footnote{radius of the shell} &
$T_\mathrm{ini}$\footnote{initial temperature of the shell} &
 composition \\
&$\mathrm{km~s^{-1}}$&$M_\odot$&$10^{15}$ cm&
$10^4$ K& \\
\hline
M01 & 4,000 & 10 & 2.4 & 1  & solar \\
M02 & 4,000 & 10 & 2.4 & 4   &  solar \\
M03 & 4,000 & 20 & 7.2 & 1.7  &  solar \\
M04 & 2,000 & 20 & 7.2 & 1.7  &  solar \\
M05 & 8,000 & 20 & 7.2 & 1.7  &  solar \\
M06 & 4,000 & 20 & 0.72 & 13   &  solar \\
M07 & 4,000 & 20 & 7.2 & 1.7   &  C 50 \% + O 50 \% \\
\hline
\end{tabular}
\end{minipage}
\end{table}

\citet[][]{smith2007} (SM07 hereafter) investigate a SLSN LC model
resulting from a circumstellar shell through which a shock wave has
gone through.
This model basically corresponds to
the former picture of the shock breakout, i.e.,
shock breakout at the surface of a dense shell.
In other words, SM07 consider a dense shell from which photons start to emit after
the passage of a shock wave.
They apply the LC model of adiabatically cooling SN ejecta
formulated by \citet{arnett1980}.
By simply comparing the shape of the model LC to the $R$-band LC
of SN 2006gy, SM07 concluded that SN 2006gy can result from a shocked
circumstellar shell.

However, there are many simplifications involved in the
SM07 model which are not discussed so far.
The existence of the recombination wave in shocked
H-rich CSM is presumed to affect the LC as in the case of Type IIP
SNe \citep[e.g.,][]{grassberg1971,falk1977,kasen2009,bersten2011}.
In addition, SM07 compare a bolometric LC obtained from their model
to the $R$-band LC of SN 2006gy.
As the shocked shell should have temperature close to $10^4$ K
at the beginning to explain the observed SN 2006gy properties \citep{smith2010},
a large fraction of emitted photons is not in the $R$ band and
the bolometric correction should be considered.

To see the importance of these neglected effects in the SM07 analytic model,
we numerically follow the system suggested to explain SN 2006gy in SM07
by using a numerical radiation hydrodynamics code \verb|STELLA|.
\verb|STELLA| is a one-dimensional radiation hydrodynamics code.
\verb|STELLA| can treat multi-frequency radiation
and the opacity is calculated based on the physical parameters
under the assumption of the local thermodynamic equilibrium.
This assumption is a good approximation to obtain a LC from a shocked shell
because of its large density and optical depth.
Thus, \verb|STELLA| is a suitable code to see the effect of
the bolometric correction and recombination which are not taken
into account in the SM07 analytic model.
For the details of \verb|STELLA|, we refer to other articles
\citep[e.g.,][]{blinnikov1993,blinnikov2000,blinnikov2006,blinnikov2011}.

We start by showing the initial conditions of the models which
are constructed based on SM07 in Section \ref{model} and
then show the numerical results in Section \ref{results}.
Discussion is presented in Section \ref{discussion} and
we conclude in Section \ref{conclusions}.
We apply the same distance to the host galaxy (73.1 Mpc) and
extinctions (Galactic $A_R=0.43$ mag + host $A_R=1.25$ mag)
as in \cite{smith2007b} to the observed LC of SN 2006gy.

\section{Models}\label{model}
Initial conditions of our numerical calculations are constructed
based on SM07 at first.
We do not follow shock propagation in shells to make the initial
conditions the same as those of SM07.
The initial conditions are supposed to result from the shock passage
in the dense circumstellar shell.
Since we do not treat the shock wave,
the smearing term in \verb|STELLA| code which
affects the conversion efficiency from kinetic energy to radiation
discussed in \citet{moriya2012b} does not affect the results
obtained in this paper.

Table \ref{table1} is the list of our initial conditions.
The initial radius of the SM07 model suggested for SN 2006gy
is $2.4\times 10^{15}$ cm.
The mass of the shocked shell in the model is $10~M_\odot$. We assume that the system is homologously expanding
and the outermost layer velocity is $4,000~\mathrm{km~s^{-1}}$, as assumed in
SM07. The initial temperature is set constant in the entire shell.
We try two temperatures
for the SM07 system, namely, $10^4$ K (Model 01 or M01) and
$4\times 10^4$ K (M02).
The composition is solar in these models.

We also investigate several configurations other than those suggested in SM07.
M03 has the same velocity but the initial radius is three times larger 
than that of the SM07 model. The temperature is set to $1.7\times 10^4$
K to match the observed luminosity of SN 2006gy.
The mass is increased to $20~M_\odot$ in M03 to keep the shell optically thick.
M04 and M05 have the same density structure as M03 but
the velocity is 0.5 and 2 times of M03, respectively.
We also show results of M06, which is more compact than the SM07 model.
The composition is solar in M03-M06. M07 has 50\% of carbon and 50\% of oxygen,
while other properties are the same as in M03.

\begin{figure*}
\begin{center}
 \includegraphics[width=\columnwidth]{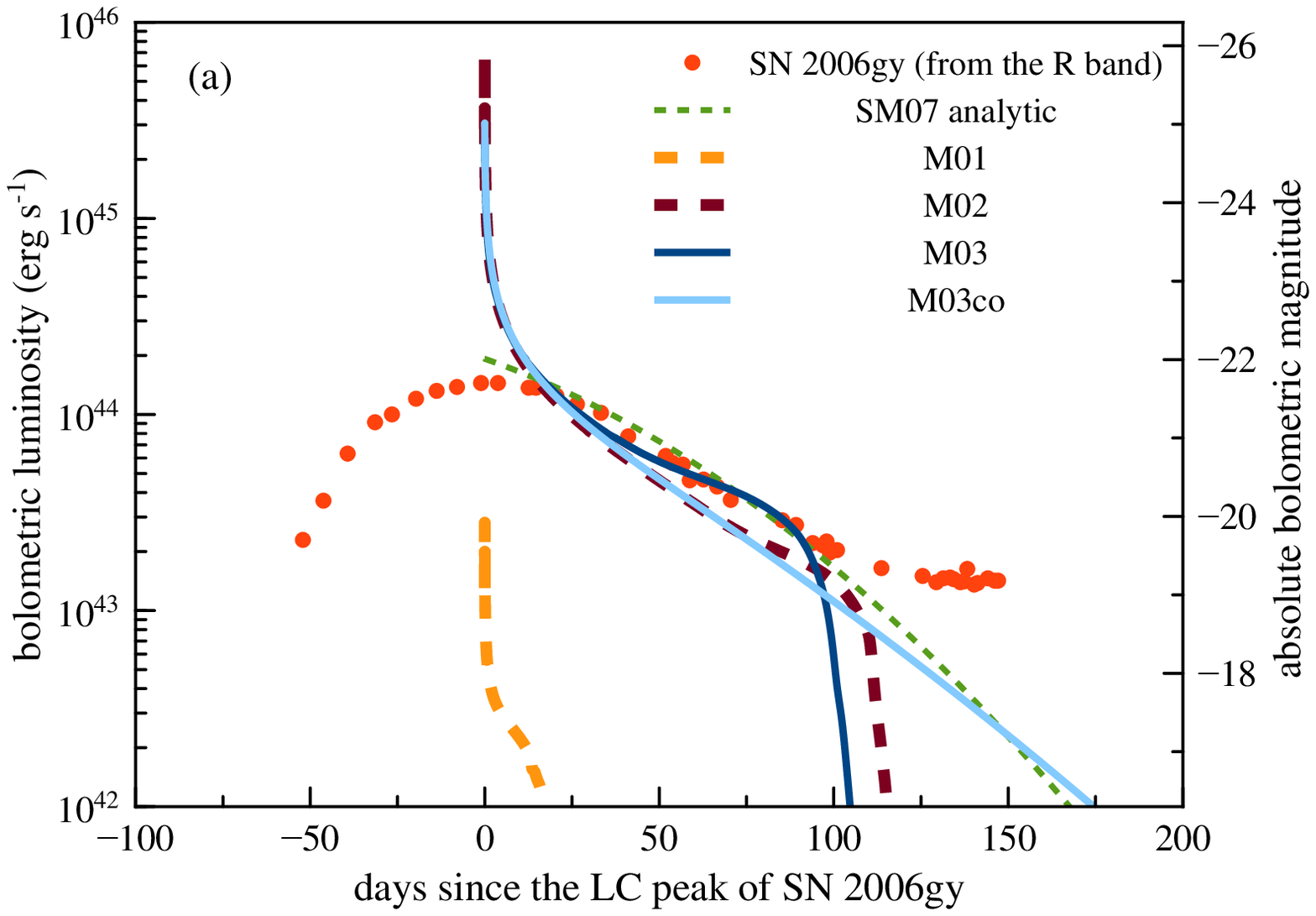}
 \includegraphics[width=0.965\columnwidth]{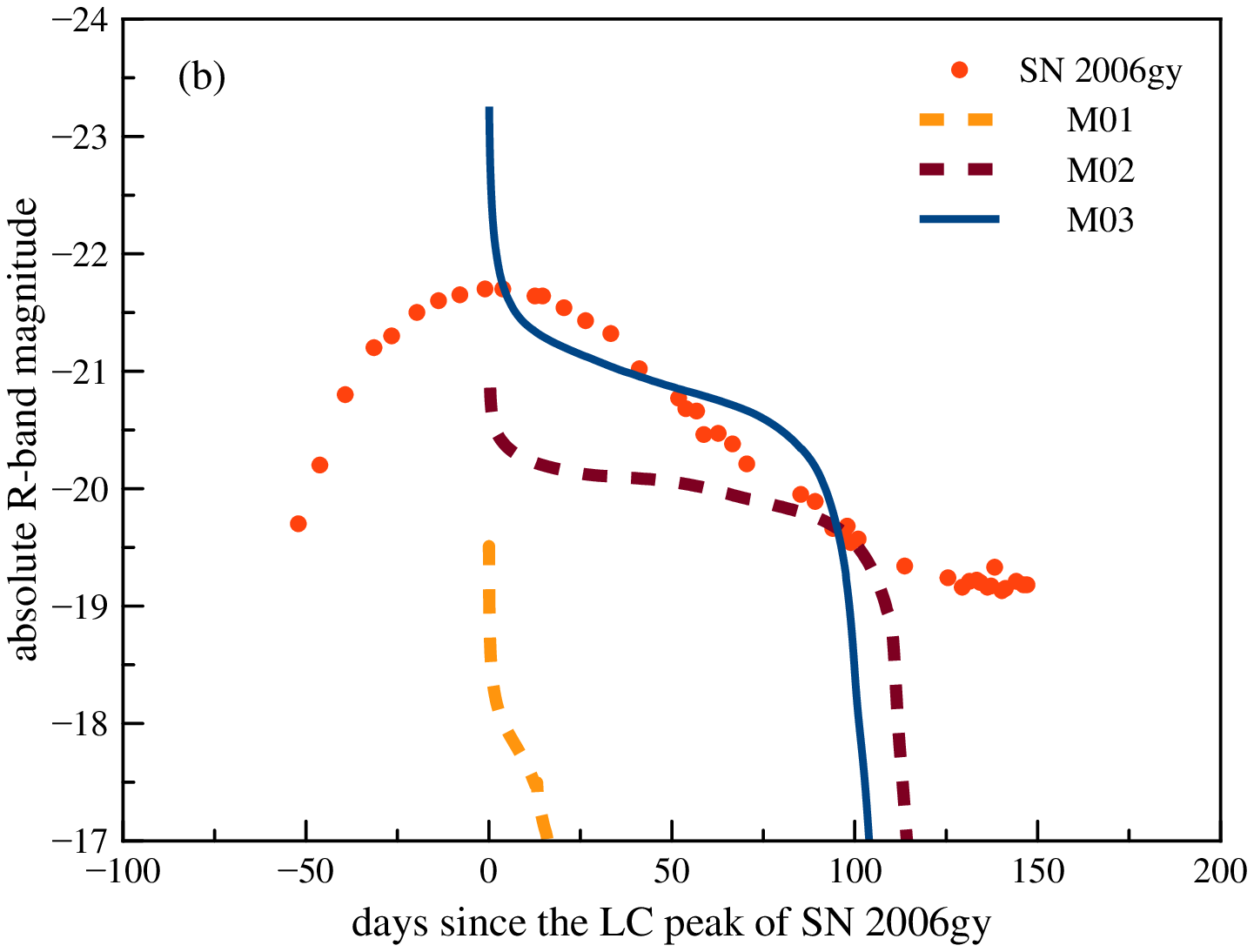}
  \caption{
Bolometric (a) and $R$-band (b) LCs of shocked shells.
}
\label{M01}
\end{center}
\end{figure*}

\section{Synthetic Light Curves}\label{results}
Figure \ref{M01}a shows the bolometric LCs of M01-M03.
At first glance, we find that no LCs are consistent
with that of SN 2006gy.
Furthermore, $R$-band LCs shown in Figure \ref{M01}b 
are found to be more different from the $R$-band LC of SN 2006gy.
This indicates the importance of the bolometric correction.
We discuss the LC behaviors in this section
but most of our discussion can be found in the previous studies,
e.g., \citet{grassberg1971,falk1977}.

The bolometric LCs start with the initial peak.
The peak bolometric luminosity is $4\pi R_o^2 \sigma T_\mathrm{ini}^4$,
where $\sigma$ is the Stephan-Boltzmann constant.
At first, the bolometric luminosity decreases due to the adiabatic expansion of the
shell.
If we assume that the homologously-expanding shell is radiation-dominated at early
phases,
the bolometric luminosity should decrease following $\propto t^{-2}$.
This rapid decrease in the bolometric luminosity appears in our
numerical models.

SM07 suggest that the bolometric LC of shocked shells
would rise following $\propto t^2$ at first because
shocked materials expand homologously and
the shocked shell is just an expanding blackbody.
However, if we take into account the decrease in the blackbody
temperature of the shell due to the adiabatic expansion
and the lack of any heat sources,
the effect of the temperature decline on the bolometric luminosity
($\propto t^{-4}$ in radiation-dominated shells)
is larger than the effect of the radius increase on the bolometric
luminosity ($\propto t^{2}$).
In fact, our synthetic LCs do not show the luminosity increase
and the luminosity just declines. 

\begin{figure}
\begin{center}
 \includegraphics[width=\columnwidth]{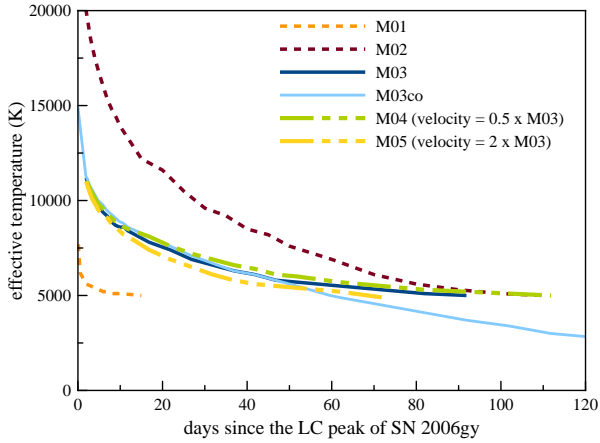}
  \caption{
Evolution of the effective temperatures.
}
\label{Tphoto}
\end{center}
\end{figure}

\begin{figure}
\begin{center}
 \includegraphics[width=\columnwidth]{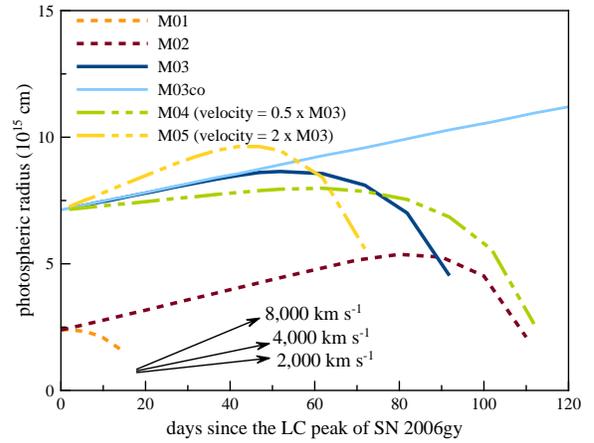}
  \caption{
Evolution of the photospheric radii.
The arrows in the figure shows the directions of the 
evolution of the radius with a given constant velocity.
}
\label{Rphoto}
\end{center}
\end{figure}

After the initial rapid luminosity decline,
bolometric LCs start to be affected by photons diffused in the shell
and begin to follow the diffusion
model of \citet{arnett1980}.
From this point, the SM07 analytic model starts to work.
M01, whose initial temperature ($10^4$ K) is close to the blackbody
temperature of SN 2006gy at the LC peak, is too faint at this epoch
to explain SN 2006gy because of the initial rapid luminosity decline
due to the adiabatic expansion.
With the initial configuration suggested by SM07,
the temperature should be around $4\times 10^4$ K (M02) to explain the
luminosity of SN 2006gy but it is inconsistent with the observed
blackbody temperature (Figure \ref{Tphoto}).
M03 has a larger radius than those of M01 and M02. Thus, the required
temperature to get the same luminosity is small ($1.7\times 10^4$ K)
and it is close to the observed values.

Although the bolometric LCs at these epochs seem to follow the observed
bolometric LC, we should be careful because the bolometric LC of SN
2006gy is obtained by the $R$-band LC without the bolometric correction.
We need to compare LCs in the $R$ band (Figure \ref{M01}b).
We find that the numerical $R$-band LCs do not match the observed $R$-band LC
even in the models which give a good fit in Figure \ref{M01}a.
This is simply because of high temperatures in the shell and most of the
emitted photons are not in the $R$-band.
We note that the strong H$\alpha$ line observed in SN 2006gy is in the
$R$-band and the direct comparison between our numerical
$R$-band LCs and the observed
$R$-band LC can be inappropriate. However, the H$\alpha$
luminosities of SN 2006gy at the epochs we are interested in
is just $\sim 10^{41}~\mathrm{erg~s^{-1}}$ \citep{smith2010} and the
bolometric correction remains to be important.

\begin{figure*}
\begin{center}
 \includegraphics[width=\columnwidth]{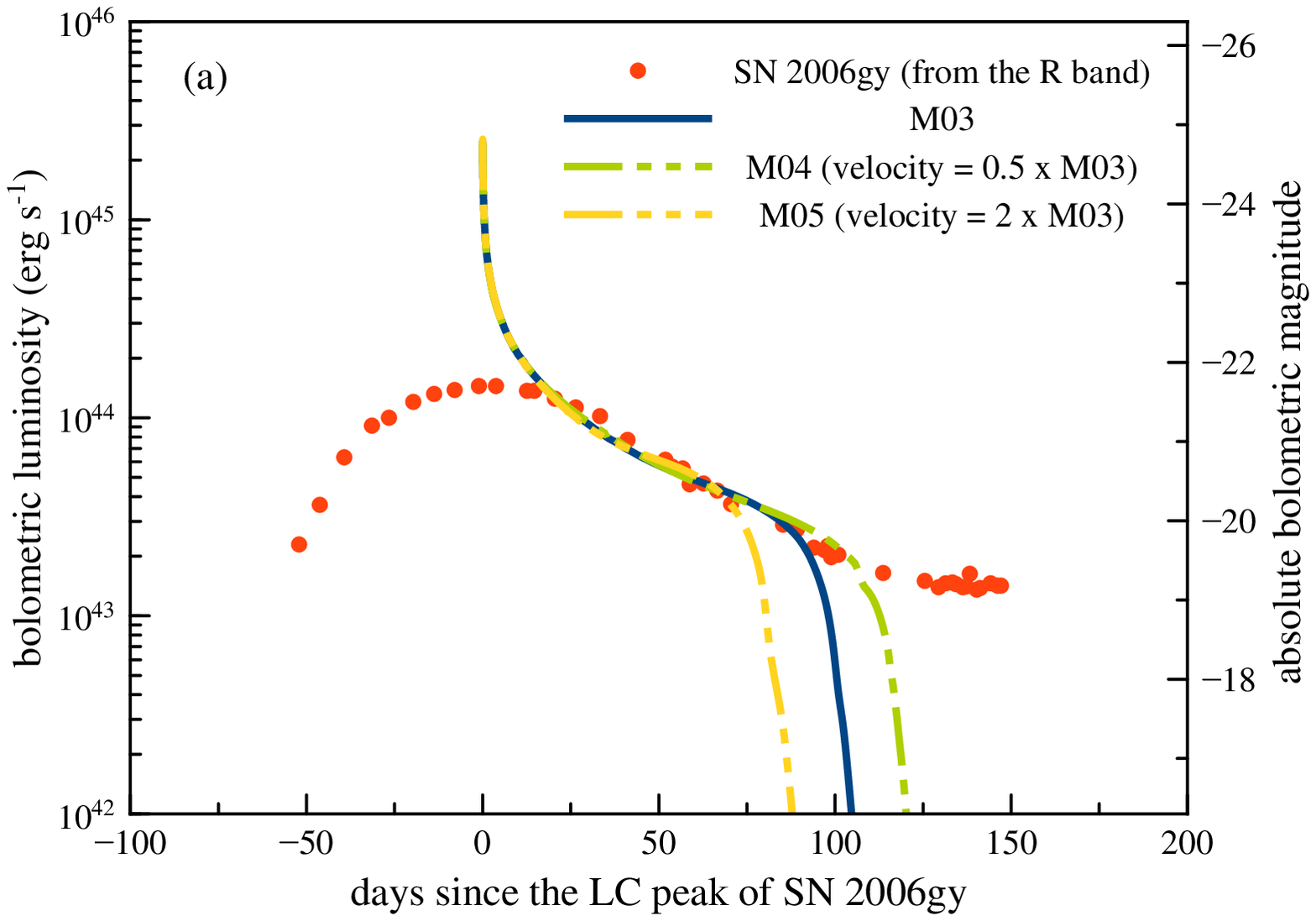}
 \includegraphics[width=0.965\columnwidth]{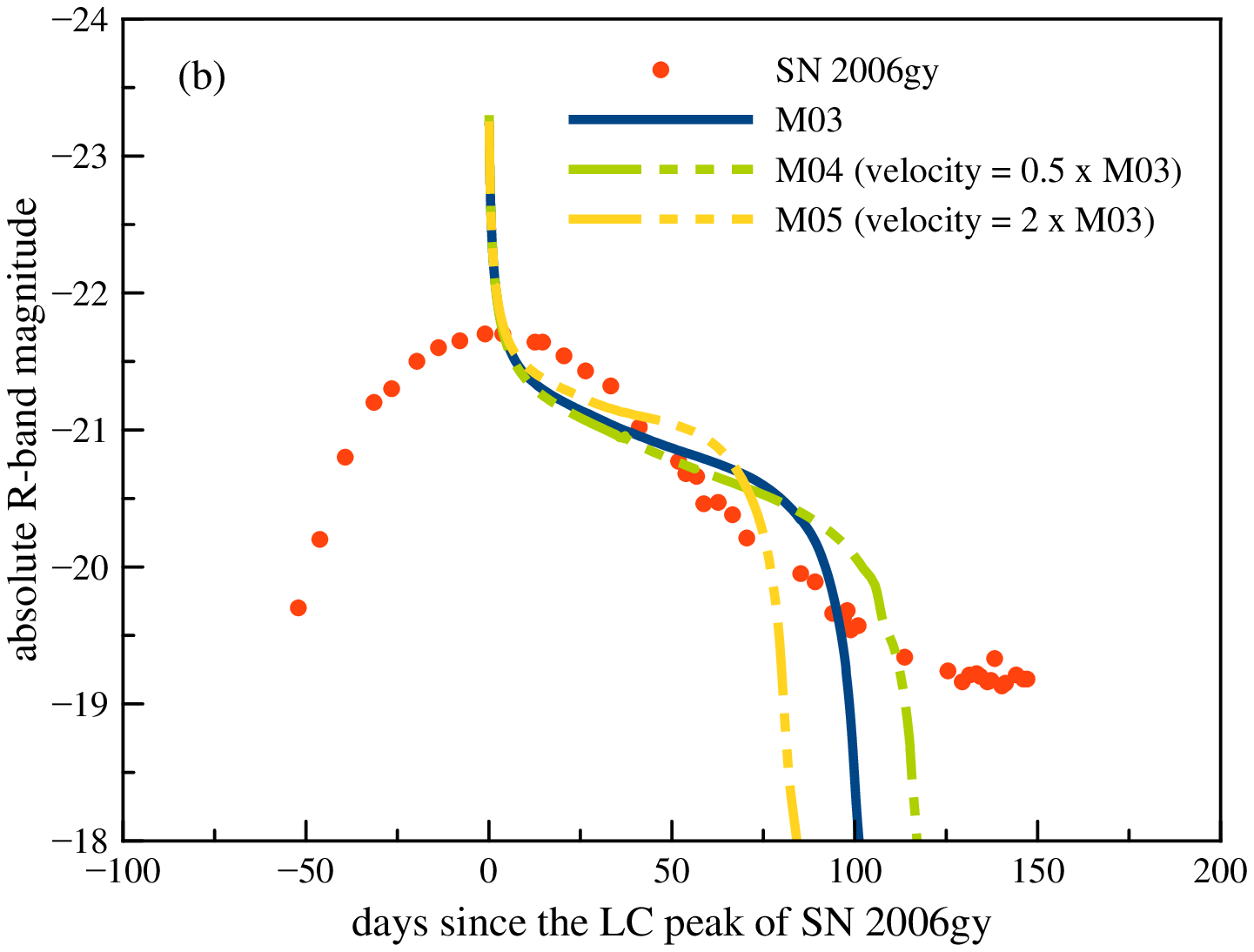}
  \caption{
Bolometric (a) and $R$-band (b) LCs of the models
with different initial velocities.
}
\label{R}
\end{center}
\end{figure*}

Another unavoidable and important consequence of the SM07 model
which is not discussed in SM07 is the existence of the recombination wave in
the shocked shell. In the SM07 model, there are no energy
sources in the shell because the shock has already passed the shell
and the shocked shell just cools down.
At one epoch, the temperature should reach the recombination
temperature as is the case for Type IIP SNe.
This is not the case for the continuous ejecta-CSM interaction models
because there remains an energy source (shock waves) which can keep the shell
ionized until the shock wave goes through the dense shell.

The effect of the recombination can be seen by comparing
M03 and M03co in Figure \ref{M01}a. M03 is calculated
with our standard opacity table which takes recombination into account.
M03co (M03 constant opacity) is calculated
by forcing the scattering opacity of the system to be
$0.34~\mathrm{cm^2~g^{-1}}$, which corresponds to
the fully ionized solar composition materials.
At first, when the shell is above the recombination temperature,
the two LCs follow almost the same track. Then, two LCs start to deviate
when the outermost layer reaches
the recombination temperature at around 40 days
since the LC peak (Figure \ref{Tphoto}).
The recombination wave, and thus the photosphere,
move inside (in Lagrangian sense) after this epoch.
They eventually reach the bottom of the shell and
the LC suddenly drops.
On the other hand, the LC with the constant opacity continues
to decline monotonically, roughly following the SM07 analytic model.
Figure \ref{Rphoto} shows the photospheric radii of the models
and the effect of the recombination is clear.

Another important consequence caused by the existence of the
recombination is the strong dependency of LCs on the shell velocity.
The epoch when the outermost layers reach the recombination temperature
and the recession velocity of the recombination wave in the shell
are affected by the shell velocity. This is simply
because adiabatic cooling becomes more efficient in faster shells.
M04 and M05 have slower and faster shells, respectively, than M03
and their LCs are presented in Figure \ref{R}.
At first, the LCs are expected to differ when the recombination
start to play a role in the shells. However, the difference
at this epoch is not significant according to our calculations.
The time of the drop in the LCs clearly differs and the faster shells
have earlier drops due to the faster recombination wave.

\begin{figure}
\begin{center}
 \includegraphics[width=\columnwidth]{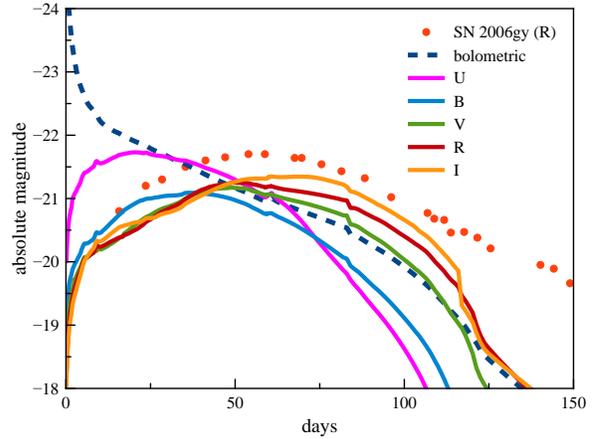}
  \caption{
Bolometric and multicolor LCs from M06.
The $R$-band LC has a rising phase similar to that of SN 2006gy
because of the initial high temperature $1.3\times 10^5$ K.
However, the high temperature is inconsistent with the SN 2006gy
observations.
}
\label{M06}
\end{center}
\end{figure}

\begin{figure*}
\begin{center}
 \includegraphics[width=\columnwidth]{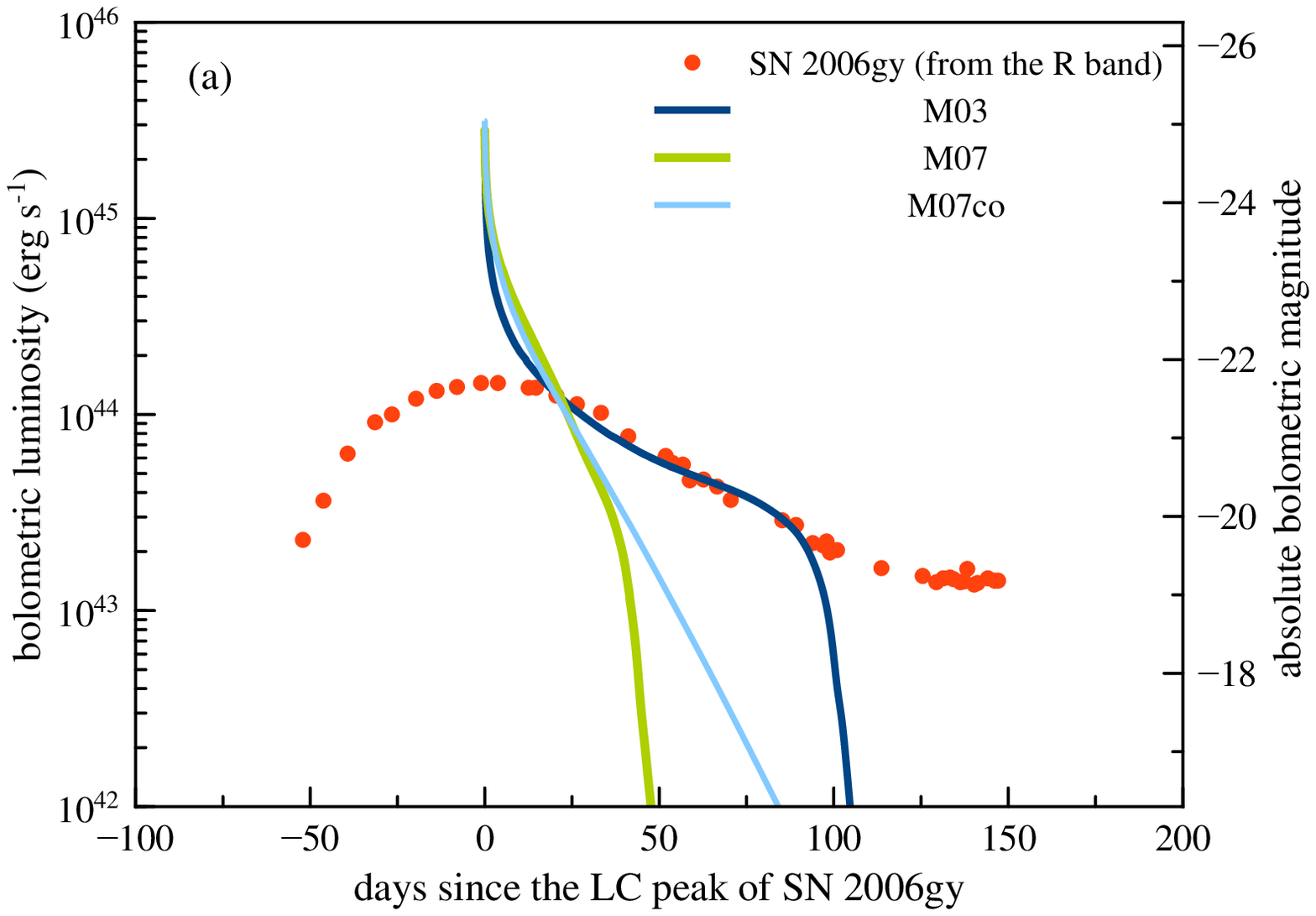}
 \includegraphics[width=0.965\columnwidth]{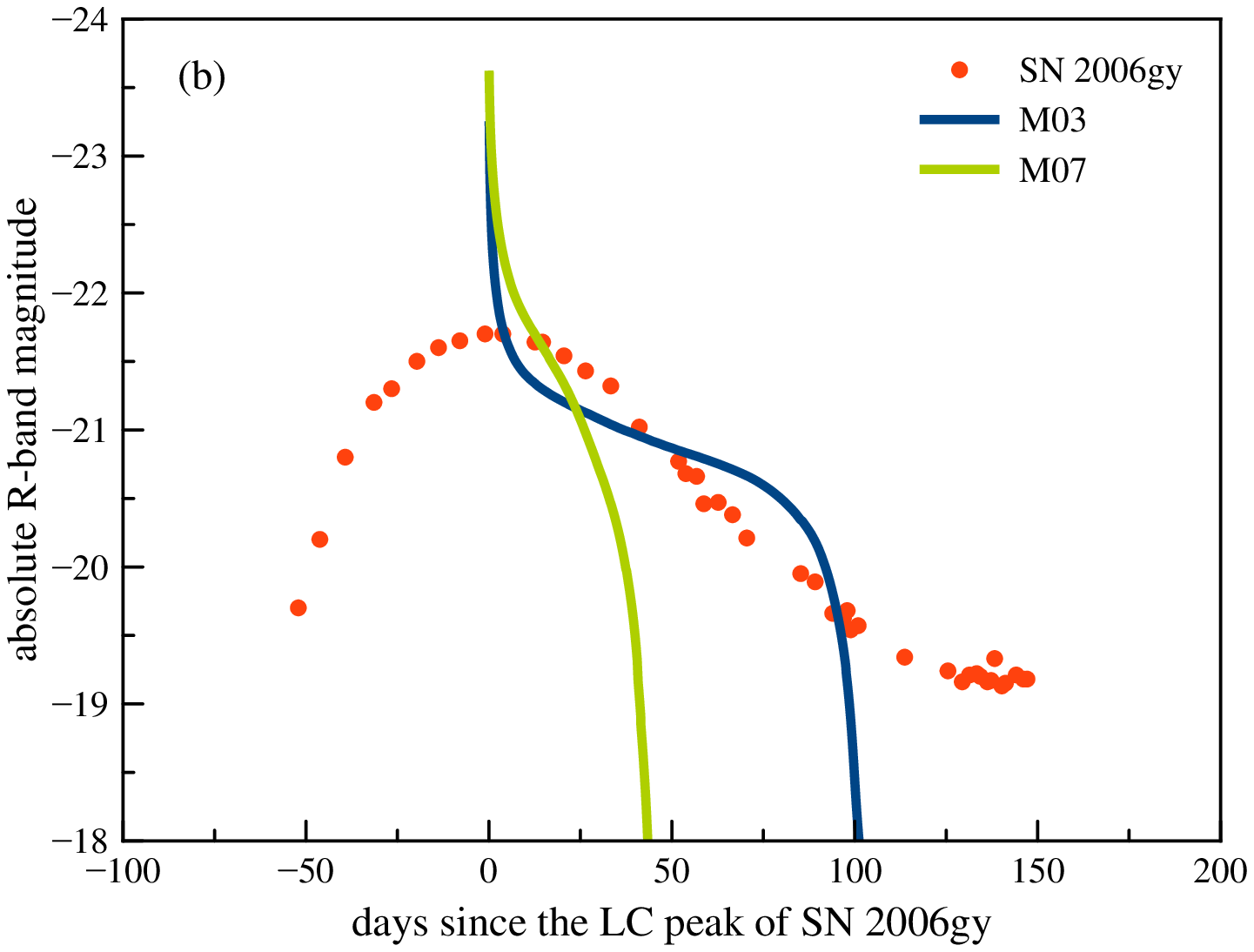}
  \caption{
Bolometric (a) and $R$-band (b) LCs from a C+O-rich shocked shell (M07).
M03 LCs are also shown for a reference.
}
\label{M07}
\end{center}
\end{figure*}

\section{Discussion}\label{discussion}

\subsection{Model with Initial R-band Luminosity Increase}
All the models we have presented so far do not have a phase
with luminosity increase and the luminosity just declines.
However, it is possible to
have a rising phase in optical LCs from a shocked shell.
Figure \ref{M06} shows one example of LC obtained from M06.
The bolometric LC and optical LCs as well as the $R$-band LC
of SN 2006gy are shown in the same figure for the illustrative purpose.
The evolution of the bolometric LC does not differ so much from the
previous models but optical LCs of M06 have a rising phase.
This is because of the initial small radius and high temperature.
The optical luminosities are low at the beginning due to the initial
high temperature. Then, as the adiabatic cooling is efficient
due to the initial small shell radius, the shell cools quickly and
optical luminosities increase accordingly.
Then, the $R$-band LC can be similar to that of SN 2006gy, although
the photospheric temperature is much higher in M06 and it is inconsistent
with the SN 2006gy observations.

\subsection{Possible Corresponding Supernovae}
LCs of shocked shells obtained by our numerical calculations
have an initial rapid decline followed by a relatively long
plateau. 
Although these features are not seen in SLSN 2006gy,
SLSN 2003ma (SLSN-II) qualitatively shows similar features \citep{rest2011}.
The LC of SN 2003ma is different from other known SLSNe.
The LC of SN 2003ma has a quick rise and quick decline 
followed by a long plateau phase which lasts for about 100 days
while other SLSNe evolves more slowly.
Then the LC drops by about 1 mag in optical and the luminosity
stays almost constant for about 1000 days after the drop.
The initial rise and decline as well as the plateau phase
which lasts for about 100 days can be seen in some synthetic optical LCs
obtained in this study (e.g., Figure \ref{M01}b),
but the plateau phase after the drop which lasts for about 1000 days
requires another emission mechanism like a continuous CSM interaction.

SN 1988Z has a similar feature to SN 2003ma, although the luminosity
is about 3 magnitude smaller \citep[e.g.,][]{turatto1993,aretxaga1999}.
Because of the LC similarity, SN 1988Z can also be related to shocked shells
\citep[see also][]{ChugaiDanziger1994}.
Depending on, e.g., radii and temperatures of shocked shells,
their luminosities can vary.
There can be many other similar SNe with variety of luminosities and
plateau durations, depending on the shell properties.

\subsection{Other Effects} 
\subsubsection{Shell of Carbon and Oxgen}
We examine a LC from a shocked shell with 50 \% of  carbon and 50\% of oxygen.
A subclass of SLSNe is known to have no hydrogen features in their
spectra and their composition is likely to be dominated
by carbon and oxygen \citep[e.g.,][]{quimby2011}.
Figure \ref{M07} shows the results. As the opacity significantly
decreases compared to the cases of the solar composition, the LC
declines much faster. In addition, due to the high recombination
temperature of carbon and oxygen, the effect of the recombination is
less significant than the cases of the solar composition.
However, we do see the significant drop in the LC due to recombination
if we compare it to the LC model (M07)
with a constant opacity ($0.1~\mathrm{cm^2~g^{-1}}$).
Thus, we should also be cautious about the applicability of SM07 model
when we try to apply it to carbon and oxygen dominated system.

\subsubsection{Opacity Table}
\begin{figure*}
\begin{center}
 \includegraphics[width=\columnwidth]{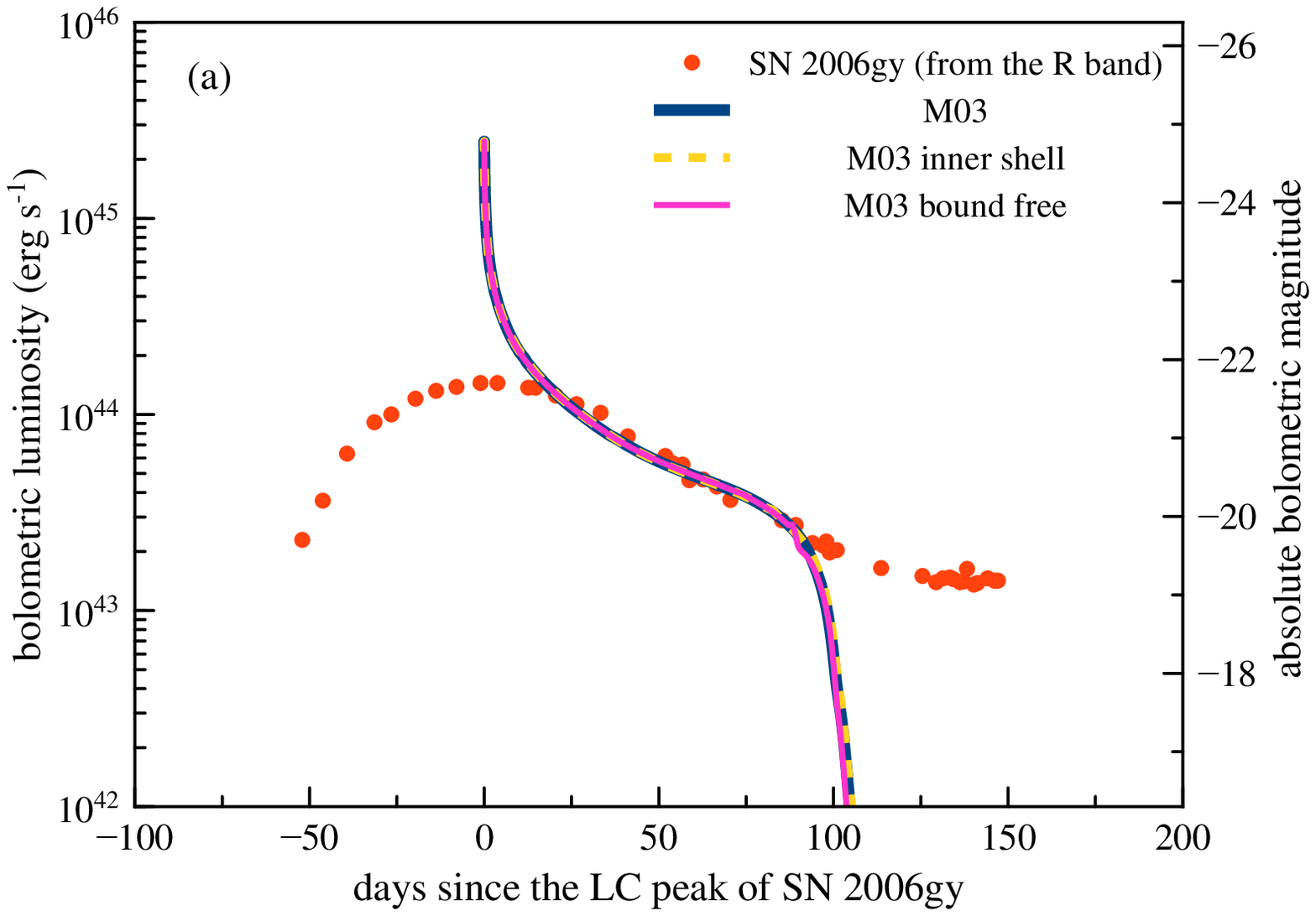}
 \includegraphics[width=0.965\columnwidth]{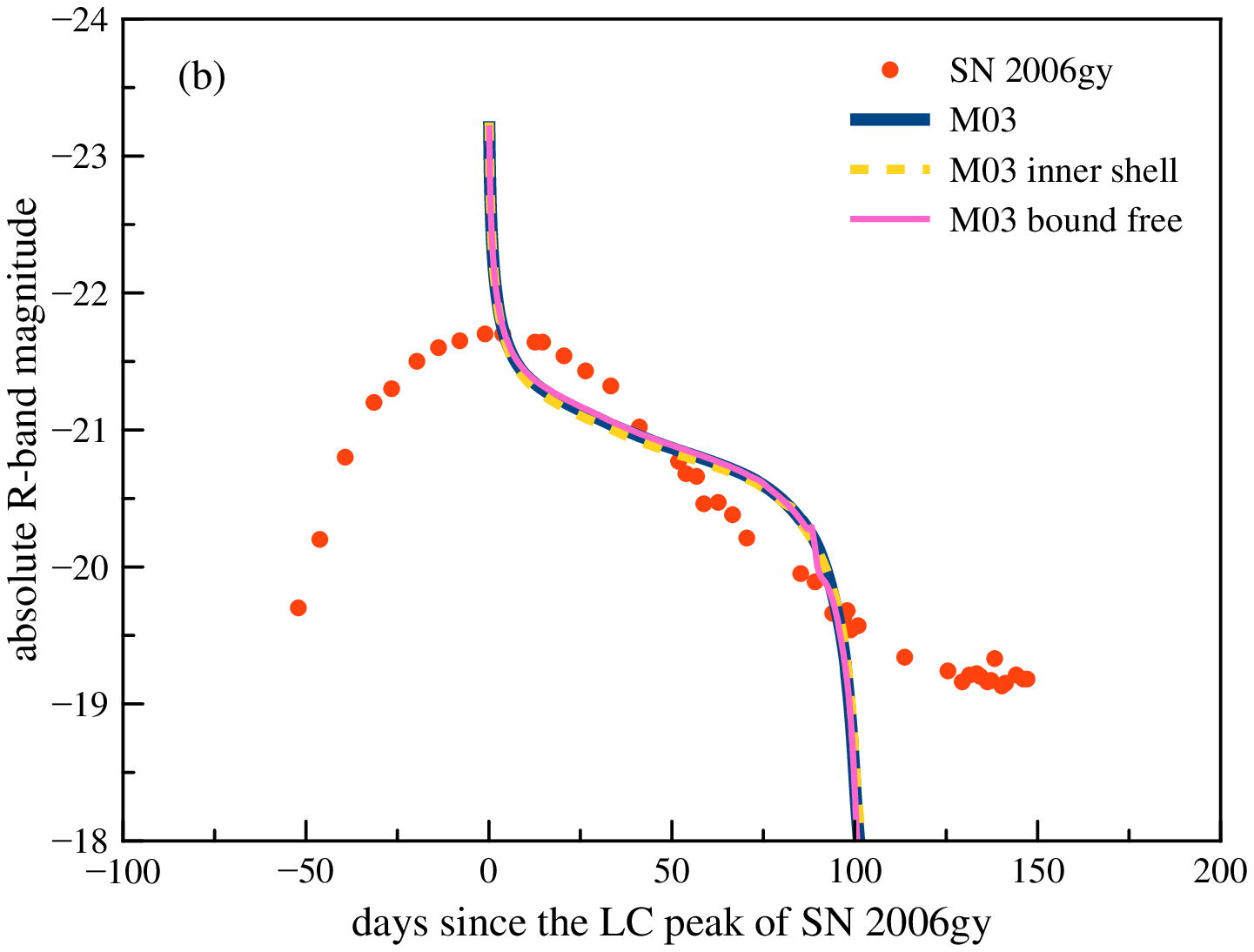}
  \caption{
Bolometric (a) and $R$-band (b) LCs with different opacity tables.
'M03' is computed with the default opacity table.
The opacity table used in 'M03 inner shell' additionally takes into
account the inner-shell photoionization.
In 'M03 bound free', some exited levels are added in bound-free
 absorptions to the default opacity table.
}
\label{opacity}
\end{center}
\end{figure*}

We also investigate the effect of the different opacity tables
adopted in \verb|STELLA| (Figure \ref{opacity}).
The detailed descriptions about the three opacity tables
adopted here can be found in \citet{tominaga2011}.

In one opacity table ('inner shell' in Figure \ref{opacity}),
the inner-shell photoionization is additionally taken into account.
The opacity table used so far in this paper
assumes that all atoms and ions,
except for hydrogen, are in the ground state
to obtain the opacity from the bound-free absorptions
as in \citet{eastman1993}.
In the other opacity table used in this section
('bound free' in Figure \ref{opacity}),
exited levels are also taken into account in bound-free absorptions.

As can be seen in Figure \ref{opacity} in which LCs from the
three opacity tables are shown,
the difference caused by the different opacity tables is small.
Thus, the effect of inner-shell photoionizations
and excited levels in bound-free absorptions
on LCs is negligible in the system
we are interested in this paper.

\section{Conclusions}\label{conclusions}
We numerically investigate LC properties of shocked shells which are
suggested to account for SN 2006gy by SM07 analytically.
We show that shocked shells fail to explain the rising part of the
SN 2006gy LC because of the adiabatic cooling.
In addition, in the declining part, we show the importance of
the effects of the bolometric correction and
the recombination which are not taken into account in SM07.
SM07 compare their analytic bolometric LCs to the observed
$R$-band LC of SN 2006gy but the high temperature of the shell
which is estimated from the spectral observations is
against this assumption and $R$-band LCs of shocked shells
becomes flatter than bolometric LCs.
Recombination also makes the LCs flatter than those analytically
suggested in SM07.

We also show the effect of the expansion speed of shocked shells.
It affects the propagation velocity of the recombination wave in the
shells and changes duration of the LCs.
The composition of the shells alters the opacity and is shown to
change the LCs very much.

Although we show that the LC of SN 2006gy is not consistent with
the shocked circumstellar shell model,
this kind of system can exist.
We suggest that SLSN 2003ma and SN 1988Z may come from
a shocked shell because of their qualitative
LC similarity to our numerical results.
In addition,
our numerical modeling indicates that shocked shells can be bright
(sometimes more than $\sim -23$ mag in optical) and
their luminosities can drop more than 1 magnitude in optical
within 1 day (M03).
This kind of object can be observed in the future transient surveys and
they can fill the bright and fast declining
part in the explosive transient phase space \citep[e.g.,][]{kulkarni2012}.
The existence of such transients from a shocked shell indicates
the existence of the explosive mass loss just before the explosions of
some kinds of stars which is currently not known well and
they can provide us a clue to understand such mass loss.

\section*{Acknowledgments}
We thank the anonymous referee for helpful comments.
This work has been done when T.J.M. and S.I.B. visited Max-Planck Institute for
Astrophysics and we acknowledge the support from the institute.
T.J.M. is supported by the Japan Society for the Promotion of Science
Research Fellowship for Young Scientists $(23\cdot5929)$.
S.I.B., P.V.B. and E.I.S. are supported partly by the grants
RFBR 10-02-00249, 10-02-01398,
by RF Sci.~Schools 3458.2010.2 and 3899.2010.2,
and by a grant IZ73Z0-128180/1 of the Swiss National Science
Foundation (SCOPES).
S.I.B., P.V.B., A.D.D. are supported by the grant
of the Government of the Russian Federation (No 11.G34.31.0047).
All the numerical calculations were carried out on the general-purpose
PC farm at Center for Computational Astrophysics, CfCA, of National
Astronomical Observatory of Japan.
This research is also supported by World Premier International Research Center Initiative, MEXT, Japan.

\label{lastpage}

\end{document}